\documentclass[aps,prl,twocolumn,nofootinbib,superscriptaddress]{revtex4}
\usepackage{epsfig}
\usepackage{bm}
\usepackage{amssymb}
\usepackage{amsmath}
\usepackage{color}
\usepackage{subfigure}
\usepackage[normalem]{ulem}

\begin{document}

\title{ The importance of the bulk viscosity of QCD in ultrarelativistic heavy-ion collisions }

\author{S.\ Ryu}
\affiliation{Department of Physics, McGill University, 3600 University
Street, Montreal,
QC, H3A 2T8, Canada}

\author{J.-F.\ Paquet}
\affiliation{Department of Physics, McGill University, 3600 University
Street, Montreal,
QC, H3A 2T8, Canada}

\author{C.\ Shen}
\affiliation{Department of Physics, McGill University, 3600 University
Street, Montreal,
QC, H3A 2T8, Canada}

\author{G.\ S.\ Denicol}
\affiliation{Department of Physics, McGill University, 3600 University
Street, Montreal,
QC, H3A 2T8, Canada}

\author{B.\ Schenke}
\affiliation{Physics Department, Brookhaven National Laboratory, Upton, NY
11973, USA}

\author{S.\ Jeon}
\affiliation{Department of Physics, McGill University, 3600 University
Street, Montreal,
QC, H3A 2T8, Canada}

\author{C.\ Gale}
\affiliation{Department of Physics, McGill University, 3600 University
Street, Montreal,
QC, H3A 2T8, Canada}

\begin{abstract}
We investigate the consequences of a nonzero bulk viscosity coefficient on the transverse momentum spectra, azimuthal momentum
anisotropy, and multiplicity  of charged hadrons produced in heavy ion collisions at LHC energies. The agreement between a 
realistic 3D hybrid simulation and the experimentally measured data considerably improves with the addition of  a bulk viscosity 
coefficient for strongly interacting matter. This paves the way for an eventual quantitative determination of several QCD transport 
coefficients from the experimental heavy ion and hadron-nucleus collision programs.
\end{abstract}

\maketitle
\date{\today }


\textit{1. Introduction}. Ultrarelativistic heavy ion collisions realized
at the Relativistic Heavy Ion Collider (RHIC) and the Large Hadron Collider
(LHC) are able to reach energies high enough to create and study the
quark-gluon plasma (QGP), a novel state of nuclear matter where the quark and gluon
degrees of freedom become manifest \cite{gale}. This hot and
dense nuclear medium was found to behave like an almost perfect fluid, with
one of the smallest shear viscosity to entropy density ratios, $\eta/s$, in nature
\cite{Gyulassy:2004zy,whitepaper1,whitepaper2,whitepaper3,whitepaper4}.
Currently, one of the main theoretical challenges in nuclear physics is to
model such collisions and extract from experiment the transport properties
of this new phase of nuclear matter.

Fluid-dynamical models have been highly successful in describing the
production of hadrons in heavy ion collisions. The azimuthal momentum
anisotropy of hadrons in particular has been shown to be a sensitive
probe of the shear viscosity of the QGP \cite{Romatschke:2007mq}, and has
been used repeatedly to estimate this transport coefficient \cite{Song:2012ua}. One limitation of this extraction procedure is the
uncertainty associated with the early time dynamics of the collisions: the
azimuthal momentum distribution of hadrons is known to be closely related
to the initial shape of the medium \cite{Alver:2010gr,Gardim:2011xv,Niemi:2012aj}. Therefore, an accurate determination 
of the shear viscosity and other transport properties of QCD matter demands further improvements in the modeling of the earliest stages of the collisions.

Recent improvements in modeling the 
early time dynamics of heavy ion collisions \cite{Schenke:2012wb,Schenke:2012fw} using the IP-Sat model of the
nucleon wavefunction \cite{IPSat} followed by a classical Yang-Mills
evolution of the gluon fields \cite{YM} led to unprecedented success
\cite{Gale:2012rq} in describing charged hadron azimuthal momentum
distributions as characterized by their harmonic coefficients $v_{n}$
($n=2,3,4$,$\cdots $). Further support for this initial state model, known as IP-Glasma, was provided
by the remarkable agreement with data of its prediction
for the event-by-event distributions of $v_{n}$ measured
by the ATLAS collaboration \cite{Aad:2013xma}. 

The same approach, however, had less success in describing the
full transverse momentum distribution of hadrons, showing clear tension with
data in the low transverse momentum region \cite{Gale:2012rq}. In this letter we show that the
inclusion of bulk viscosity, which was neglected in previous studies, can
relieve this tension. In principle, the bulk viscosity of QCD matter should not be zero for the temperatures achieved
at the RHIC and the LHC and it may become large enough to affect the evolution of the medium. In fact, simulations of heavy ion
collisions that include the effect of bulk viscosity have already been
performed \cite{Torrieri:2008ip,Rajagopal:2009yw,Bozek:2009dw,Bozek:2011ua,Bozek:2012qs,Monnai:2009ad,Denicol:2009am,
Song:2009rh,Dusling:2011fd,Noronha-Hostler:2013gga,Noronha-Hostler:2014dqa} 
and demonstrated that bulk viscosity can have a non-negligible effect on 
heavy ion observables. 

In addition to the early time description of the collision provided by the
IP-Glasma model, our calculations include
a phase of hadronic re-scatterings after the hydrodynamic evolution, implemented using the ultra relativistic quantum molecular dynamics simulation UrQMD \cite{UrQMD1,UrQMD2}. Moreover, the intermediate fluid-dynamical evolution is resolved using a more complete version \cite{Denicol:2010xn} of Israel-Stewart theory \cite{IS} that takes into account all the second order terms that couple the
shear-stress tensor and bulk viscous pressure. This hybrid approach with IP-Glasma initial conditions is found to be capable of describing simultaneously 
the multiplicity and average transverse momentum of pions, kaons, and protons when a finite bulk viscosity, of the order $\zeta/s \approx 0.3$, is included
near the QCD phase transition region. Such a finite bulk viscosity also considerably reduces, by almost 50\%, the value of the
shear viscosity needed to describe the harmonic flow coefficients.


\textit{2. Model}. 
The initial state of the medium is determined using the IP-Glasma model
with the thermalization time set to $\tau _{0} $ = 0.4 fm. 
The system then evolves following the conservation law
$\partial _{\mu }T^{\mu \nu }=0$, where  the stress-energy tensor
$T^{\mu \nu }$ is composed of the ideal part 
$T^{\mu\nu}_{\rm id} =\varepsilon u^{\mu }u^{\nu }
-\Delta ^{\mu \nu }P_{0}(\varepsilon)$
and the dissipative part
$T_{\rm diss}^{\mu\nu} = \pi ^{\mu \nu } -\Delta ^{\mu \nu }\Pi $.
Here $ \varepsilon $ is the local energy density, 
$P_{0}(\varepsilon)$ is the thermodynamic pressure according to the equation
of state, $u^{\mu }$ the fluid velocity, $ \Pi $ the bulk viscous pressure, and $\pi ^{\mu \nu }$ the shear-stress
tensor. We further introduced the projection operator 
$\Delta ^{\mu \nu }=g^{\mu \nu }-u^{\mu }u^{\nu }$ 
onto the 3-D space orthogonal to the fluid velocity. 
The equation of state, $P_{0}(\varepsilon )$, is the chemical equilibrium
one taken from Ref.~\cite{Huovinen:2009yb}.
It is a parametrization of a lattice QCD calculation
matched onto a hadron resonance gas calculation at lower
temperatures. We assume that the baryon number density and diffusion are
zero at all space-time points and our metric convention is
$g^{\mu \nu }=\mathrm{diag}(1,-1-1-1)$.

The time-evolution equations satisfied by $\Pi $ and $\pi ^{\mu \nu }$ are
relaxation-type equations derived from kinetic theory
\cite{Denicol:2012cn,Denicol:2014vaa}. These are solved numerically within the
\textsc{music} hydrodynamics simulation \cite{MUSIC,KTscheme,Marrochio:2013wla}.
Explicitly, we solve
\begin{align}
\tau_{\Pi }\dot{\Pi}+\Pi = -\zeta \theta -\delta _{\Pi \Pi }\Pi \theta
+\lambda _{\Pi \pi }\pi ^{\mu \nu }\sigma _{\mu \nu }\;,  \label{intro_1}
\\
\tau_{\pi }\dot{\pi}^{\left\langle \mu \nu \right\rangle }+\pi ^{\mu \nu }
= 2\eta \sigma ^{\mu \nu }-\delta _{\pi \pi }\pi ^{\mu \nu }\theta
+\varphi
_{7}\pi _{\alpha }^{\left\langle \mu \right. }\pi ^{\left. \nu
\right\rangle
\alpha } \notag \\
-\tau _{\pi \pi }\pi _{\alpha }^{\left\langle \mu \right. }\sigma
^{\left. \nu \right\rangle \alpha }+\lambda _{\pi \Pi }\Pi \sigma ^{\mu
\nu
}.  \label{intro_2}
\end{align}%
The above equations include all the nonlinear terms that couple
bulk viscous pressure and shear-stress tensor and have recently
been shown to be in good agreement with solutions of the 0+1
Anderson-Witting equation in the massive limit \cite{Denicol:2014mca} and
of the 1+1 Anderson-Witting equation in the massless limit
\cite{Denicol:2014xca,Denicol:2014tha}. For the sake of simplicity, the
transport coefficients $\tau _{\Pi }$, 
$\delta _{\Pi \Pi }$, $\lambda _{\Pi \pi }$, $\tau _{\pi }$, $\eta $, $%
\delta _{\pi \pi }$, $\varphi _{7}$, $\tau _{\pi \pi }$, and $\lambda
_{\pi\Pi }$ are fixed using formulas derived from the Boltzmann equation near
the conformal limit \cite{Denicol:2014vaa}. The shear viscosity coefficient is assumed to be proportional to the entropy
density, i.e., $\eta\propto s$. The bulk viscosity coefficient employed
is the same one introduced in Ref.~\cite{Denicol:2009am}, which corresponds to a parametrization
of calculations from Ref.\ \cite{Karsch:2007jc} for the QGP phase and Ref. \cite{NoronhaHostler:2008ju} for the hadronic phase. 
These two calculations are matched at $T_c=180$ MeV and the value of
$\zeta /s$ at this temperature is $\zeta /s(T_c)\approx 0.3$. This parametrization is plotted in Fig.~1 as the blue solid curve. The results shown
in this letter have a small sensitivity on the value of $\zeta /s$ near
the matching temperature, which can be doubled without leading to major modifications in our results. 

\begin{figure}[th]
\includegraphics[width=0.45\textwidth]{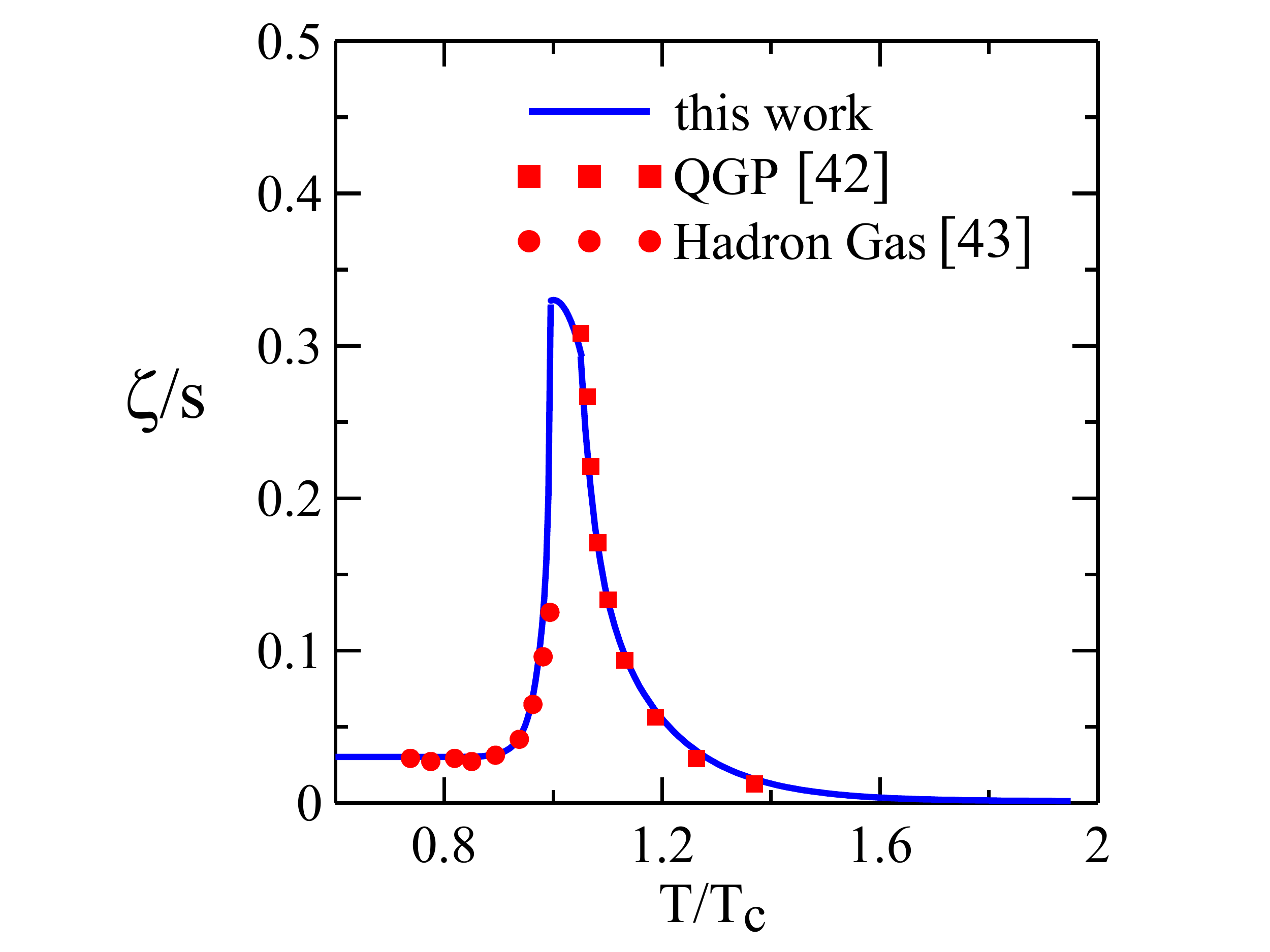}
\caption{(Color online) The bulk viscosity over entropy density parametrization used in our simulations as a function of $T/T_c$.
}
\label{fig0}
\end{figure}

At an isothermal hypersurface specified by the switching temperature 
$T_{\rm switch}$, the simulation switches from a fluid-dynamical description to a transport
description \cite{Bass:2000ib}, modeled using the UrQMD simulation. The momentum distribution of hadrons at each hypersurface element is calculated via the usual
Cooper-Frye formalism \cite{Cooper:1974mv}. The multiplicity of each
hadron species is sampled assuming that every fluid element is a
grand-canonical ensemble while the momentum of each hadron is obtained by
sampling the momentum distribution using the rejection method. We
note that the Cooper-Frye formalism requires as an input the nonequilibrium
momentum distribution of each hadron inside the fluid elements. For the
correction related to bulk viscous pressure, we employ the distribution
derived from the Boltzmann equation using the relaxation time
approximation, as described in Ref.~\cite{Rose:2014fba}. For the
shear-stress tensor nonequilibrium correction, we employ the usual \textit{ansatz} obtained
from the 14-moment approximation \cite{Monnai:2009ad,Teaney:2003kp}. The details of
how UrQMD is matched to MUSIC will be presented in an upcoming paper.  

We emphasize that the nonequilibrium corrections to the momentum distribution of hadrons at the moment of switching are still not completely understood from a theoretical point of view and represent a source of uncertainty in simulations of heavy ion collisions. However, the differential observables carry most of these uncertainties since they are more sensitive to the details of how the momentum of hadrons is distributed when converting from a hydrodynamic to a transport description. For this reason, we fix all the free parameters of our model using $p_T$--integrated observables.


\begin{figure*}[th]
\includegraphics[width=0.325\textwidth]{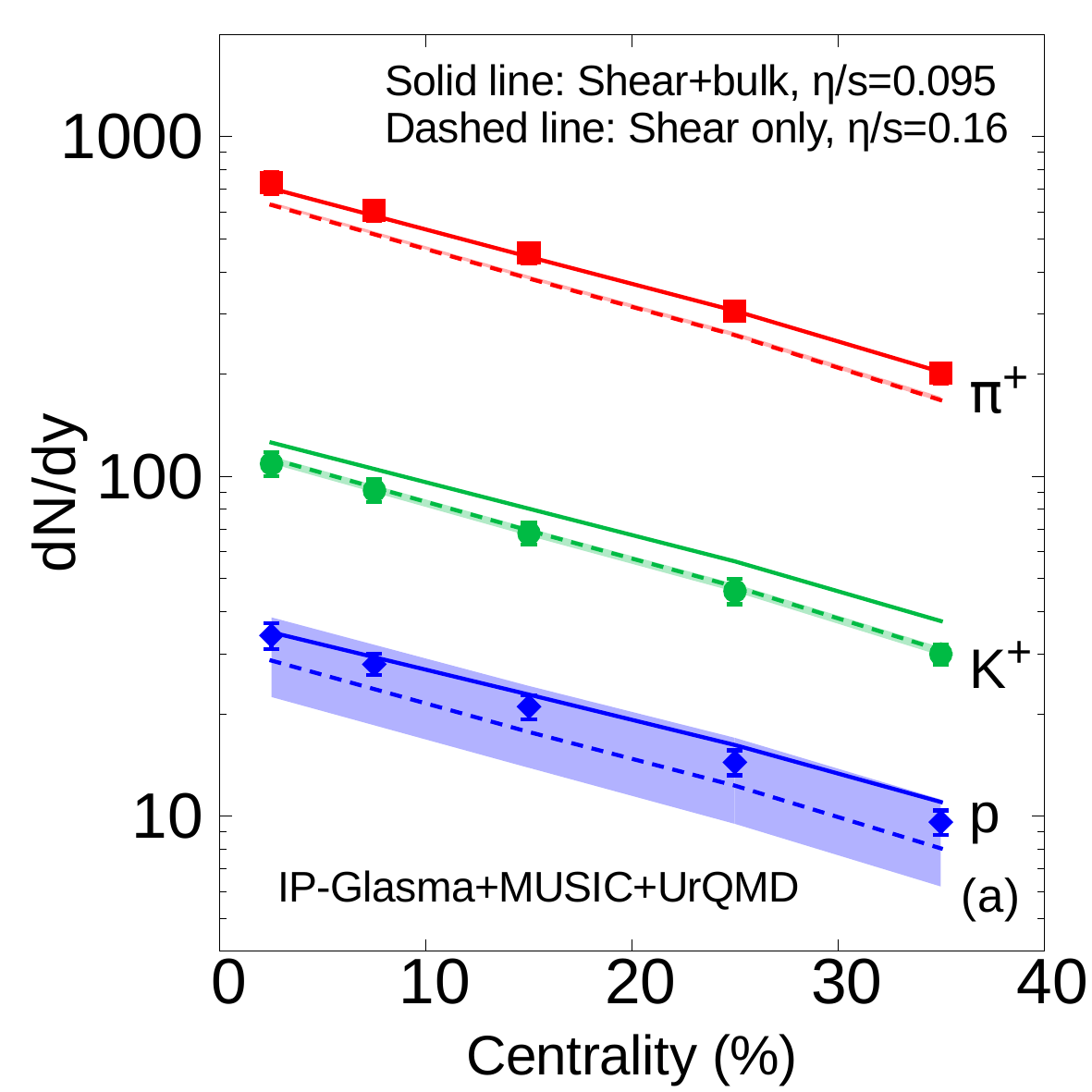}
\includegraphics[width=0.325\textwidth]{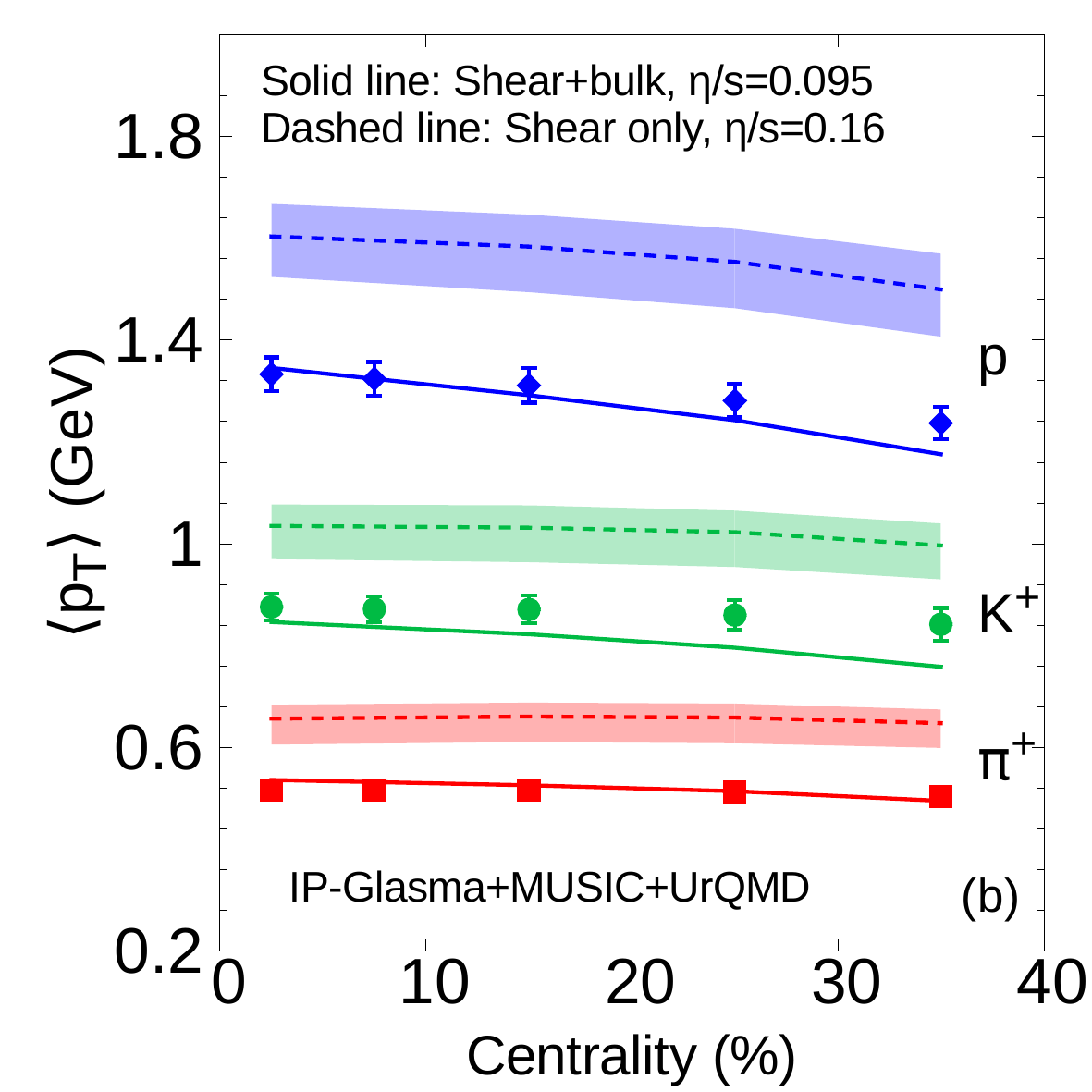}
\includegraphics[width=0.325\textwidth]{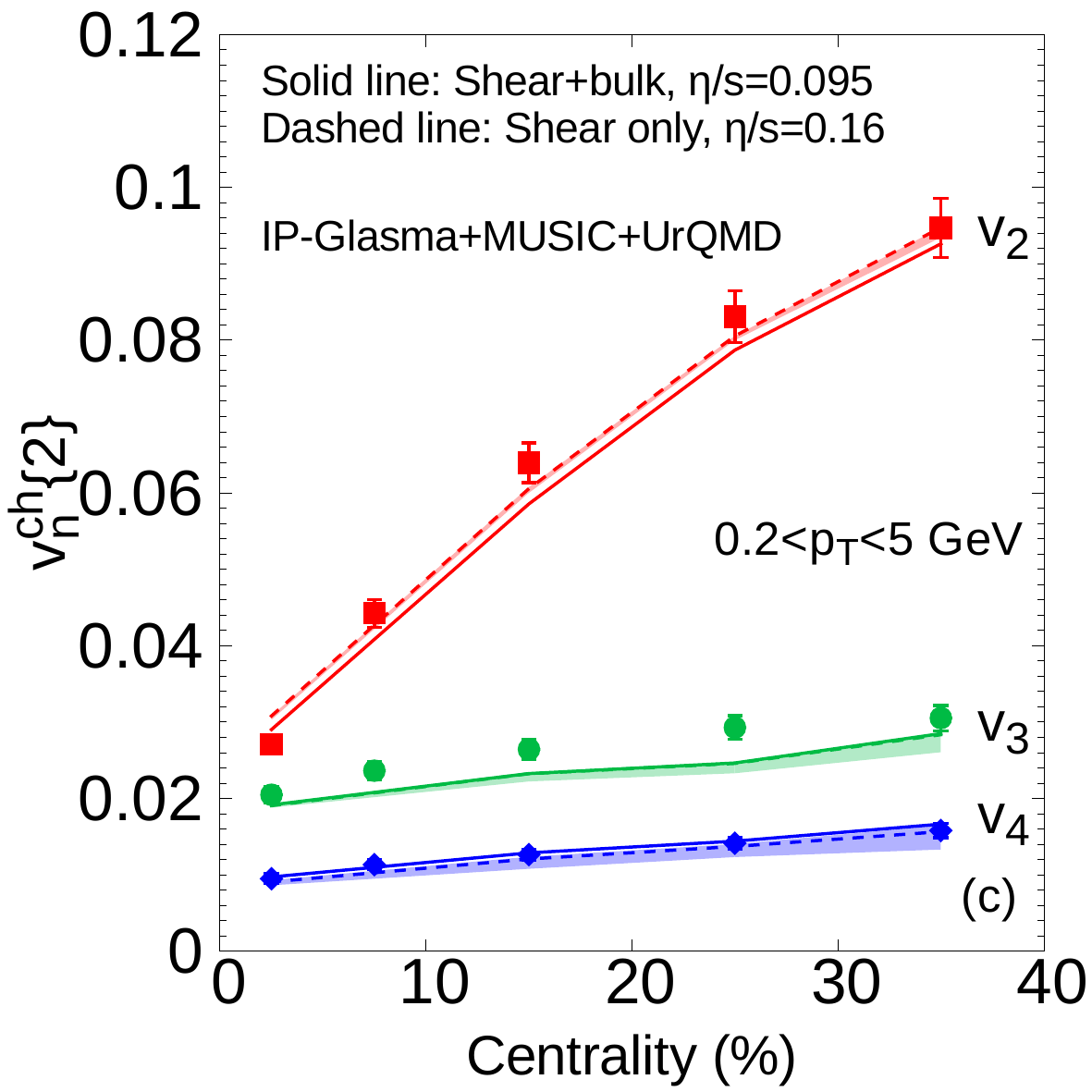}
\caption{(Color online) Multiplicity (a), average transverse momentum (b),
and flow harmonic coefficients (c) as a function of centrality. The bands around the dashed lines
show the effect of $T_{\mathrm{switch}}$ on the observables. The points correspond to measurements
by the ALICE collaboration \cite{ALICE:2011ab,Abelev:2013vea}, with bars denoting the experimental uncertainty. }
\label{fig1}
\end{figure*}

\textit{3. Results and Discussion}. In our simulations, the value of the shear viscosity coefficient is adjusted to provide a good agreement with the integrated flow harmonic coefficients, $v_n$, up to $n=4$. For the simulations that include both bulk and shear viscosity, this procedure led to the value $\eta/s=0.095$. For the simulations which include only the shear viscosity, our baseline calculation is carried out with $\eta /s=0.16$.  The larger value of $\eta /s$ compensates the reduction of momentum anisotropy due to the effect of the bulk viscosity.

The pion and kaon multiplicity, $v_n$, and, to a lesser extent, their average $p_T$, are only mildly sensitive to the choice of switching temperature between the hydrodynamic and UrQMD phases. Proton observables, on the other hand, do depend significantly on the choice of $T_{\mathrm{switch}}$. The switching temperature used in the following calculations is fixed such that a good description of the proton multiplicity and average $p_T$ is achieved for the simulation with both shear and bulk viscosities. This value is $T_{\mathrm{switch}}=145$~MeV. 

In Figs.~2 (a), (b), and (c), we show the multiplicity, average transverse
momentum of pions, kaons, and protons, and the integrated flow harmonics of charged hadrons, as a function of centrality class.
The $v_{n}\{2\}$ coefficients are calculated following the
cumulant method \cite{Bilandzic:2010jr} using the same $p_{T}$ cuts employed by the ALICE collaboration \cite{ALICE:2011ab}.
The multiplicity and average transverse momentum are
calculated without a lower $p_{T}$ cut \cite{Abelev:2013vea}.
All resonances and hadrons included in UrQMD are considered in our analyses and we neglect 
all weak decays. The solid curves correspond to the simulations that
include bulk and shear viscosities, while the dashed lines correspond to the
calculations with only the shear viscosity. The band around the dashed curves shows how the
results are modified when $T_{\mathrm{switch}}$
is varied from 135 MeV to 165 MeV. For $\left\langle p_{T}\right\rangle $ and $v_n$, the upper section of the band corresponds to the calculations with the lowest $T_{\mathrm{switch}}$ while for multiplicity it corresponds to ones with the highest $T_{\mathrm{switch}}$. The points correspond to measurements
by the ALICE collaboration \cite{ALICE:2011ab,Abelev:2013vea}.

As expected, the simulations without bulk viscosity are still able to well
describe the centrality dependence of the flow harmonic coefficients
$v_{2,3,4}\{2\}$. However, these calculations overestimate the
$\left\langle p_{T}\right\rangle $ of pions, kaons, and protons by almost 30\%. This
happens because the IP-Glasma model gives rise to an initial state with
large gradients of pressure and the subsequent fluid-dynamic expansion 
accordingly produces a significant radial flow. Therefore, in order to describe the data 
the transverse momentum of produced particles must be considerably reduced.

Including hadronic re-scatterings by itself does not reduce the $\left\langle
p_{T}\right\rangle$, modifying mostly the intermediate $p_{T}$ region of the pion spectra \cite{Song:2010aq,Song:2013qma}. Moreover, we can
see from the bands around the dashed lines in Fig.~2 that increasing the
switching temperature will not help fixing the multiplicity of pions, and is
not enough to reproduce the correct values of $\left\langle
p_{T}\right\rangle $. Finally, reducing $\eta/s$ alone 
not only is unable to sufficiently suppress the $\left\langle p_{T}\right\rangle$, but also ends up 
destroying the good description of the flow harmonic coefficients.

Including bulk viscosity leads to a suppression of $\left\langle
p_{T}\right\rangle$ and can improve our description of the data. This is because the bulk viscous pressure acts
as a resistance to the expansion or compression of the fluid. In heavy ion collisions, the expansion rate is mostly large
and positive, leading to a bulk viscous pressure that reduces the effective pressure of the system and, consequently,
slows down the acceleration of the fluid. 

As shown in Fig.~2, the calculations with bulk viscous
pressure are indeed able to provide a good description of all the $p_{T}$%
--integrated observables. The  calculated average transverse momentum of pions, kaons,
and protons are within the error bars of the ALICE measurements
\cite{Abelev:2013vea} for most of the centrality classes considered. The pion and proton multiplicities measured by
ALICE \cite{Abelev:2013vea} are well described by the model, which however systematically over-predicts
the multiplicity of kaons by $\sim 10\%$. Finally, we see that
the inclusion of bulk viscosity does not spoil the description of the flow
harmonic coefficients $v_{2,3,4}\{2\}$ as a function of centrality.
We note that the bulk viscosity reduces $v_{2,3,4}\{2\}$ by more than 10\% but this effect is 
compensated by decreasing the shear viscosity over entropy density ratio from $\eta /s=0.16$ to $\eta /s=0.095$,  
leading to a very similar quality of description. Within this study, the inclusion of bulk viscosity can therefore 
reduce the value of shear viscosity extracted from data by almost 50\%.

\begin{figure*}
\includegraphics[width=15cm]{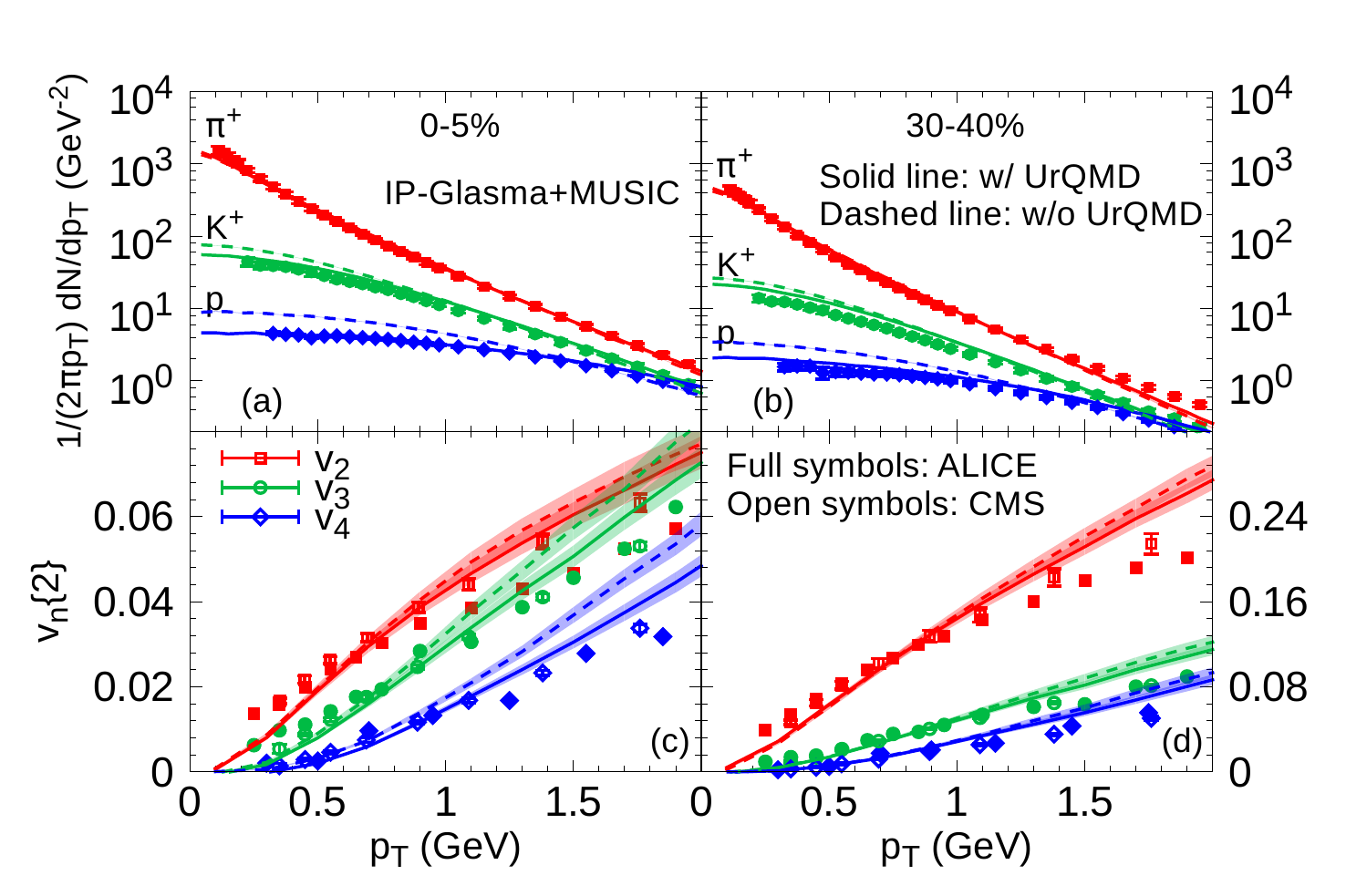}
\caption{(Color online) Transverse momentum spectra (upper panels) of
pions, kaons, and protons and harmonic flow coefficients (lower panels) as
a function of the transverse momentum. Two centrality classes are
considered: 0--5\% (left panels) and 30--40\% (right panels). The bands denote the statistical uncertainty of the calculation. 
The full and open symbols correspond to measurements by the ALICE \cite{ALICE:2011ab} and CMS \cite{Chatrchyan:2012ta,Chatrchyan:2013kba} collaboration respectively, with bars denoting the experimental uncertainty.}
\label{fig:2}
\end{figure*}

We now study $p_{T}$--differential
observables within the best fit configuration including shear and bulk viscosities.
Figure 3 shows the $p_{T}$--spectra of pions,
kaons, and protons and $v_{2,3,4}\{2\}(p_{T})$ of charged hadrons for the 0--5$\%$ and
30--40$\%$ centrality classes. The solid lines
correspond to the calculations with bulk and shear viscosity discussed
above while the dashed lines correspond to the same calculations without
the effect of hadronic re-scatterings. Note that the $p_{T}$--spectra
display reasonable agreement with the data which is in line with the
good description of the multiplicity and $\left\langle p_{T}\right\rangle $
of pions, kaons, and protons, displayed in Fig.~2. The $v_{n}\{2\}(p_{T})$
of charged hadrons shows more deviations from data, in particular the ALICE
data~\cite{ALICE:2011ab}, which is systematically smaller than the CMS
measurement \cite{Chatrchyan:2012ta,Chatrchyan:2013kba} at high $p_T$. 

We find that hadronic re-scatterings have an almost negligible effect on
pion spectra (the difference between the red dashed curve and the
solid one is barely visible in the plot) and only affects the differential
flow harmonics of charged hadrons at high $p_{T}$. On the other hand, they
play an important role in the description of kaon and, especially, proton spectra.
Without taking into account all of these effects, it would not be possible to
globally describe these observables. These findings are consistent with those from
Refs.~\cite{Song:2010aq,Song:2013qma}.

\textit{4. Conclusions}. In this letter, we discussed the effect of bulk
viscous pressure on multiplicity, average transverse momentum,
and azimuthal momentum anisotropy of charged hadrons using a
state-of-the-art simulation of ultrarelativistic heavy ion collisions. It includes  IP-Glasma initial conditions, which in combination with hydrodynamics are known 
to provide a good description of the flow harmonic coefficients, and UrQMD, which 
models the hadronic re-scatterings that follow the fluid-dynamical evolution
of the system. This fluid-dynamical evolution also considers several non-linear terms absent from several previous studies. The inclusion of bulk viscosity was found to have a large effect on the average transverse momentum of charged hadrons and on the elliptic flow coefficient. In fact, when using
the IP-Glasma initial conditions, the bulk viscosity is
essential to describe the  $p_{T}$--spectra of charged hadrons, and  leads to a considerably better
description of the data. A similar quality of description involving only shear viscosity could not be obtained in our current model. 

This work constitutes the first phenomenological investigation which shows that the bulk viscosity of QCD matter is not small, at least around the phase transition region. Our calculations suggest that $\zeta/s \approx 0.3$ or larger around $T_c$. We also showed that the inclusion of bulk viscosity considerably modifies the optimum value of shear viscosity required to describe the data, reducing it by almost 50\%. Therefore, the effects of bulk viscosity can not be neglected when extracting any transport coefficient from the data. The effects of bulk viscosity on ultracentral collisions, already briefly investigated in Ref.\ \cite{Rose:2014fba}, and on several other experimental observables will be the subject of future studies.

\textit{Acknowledgments}. The authors thank M.~Luzum, J.~B.~Rose, P.~Huovinen,
J.~Noronha, R.~Snellings, and A.~Kalweit for useful discussions. This work
was supported in part by the Natural Sciences and Engineering
Research Council of Canada, and by the U.~S.~DOE Contract
No.~DE-SC0012704. G.~S.~Denicol acknowledges support through a Banting
Fellowship of the Natural Sciences and Engineering Research Council of Canada.
B.~Schenke acknowledges support from a DOE Office of Science Early Career Award.
Computations were made on the Guillimin supercomputer at McGill University, managed by
Calcul Qu\'ebec and Compute Canada. The operation of this supercomputer is
funded by the Canada Foundation for Innovation (CFI), Minist\`ere de
l'\'Economie, de l'Innovation et des Exportations du Qu\'ebec (MEIE), RMGA
and the Fonds de recherche du Qu\'ebec - Nature et technologies (FRQ-NT).

\end{document}